\documentstyle[prl,aps,preprint,tighten]{revtex}
\begin{document}
\draft
\title{
Parity Effects on Electron Tunneling
 through Small Superconducting Islands\\
}
 \author{Gerd Sch\"on$^a$ and Andrei D. Zaikin$^{a,b}$}

\address{
 $a)$ Institut f\"ur Theoretische
 Festk\"orperphysik, Universit\"at Karlsruhe, \\
 76128 Karlsruhe,  FRG\\
 $b)$ I.E. Tamm Department of Theoretical Physics, P.N. Lebedev Physics
Institute \\
     Leninsky Prospect 53, Moscow 117924, Russia
}
\date{\today}
\maketitle
 \begin{abstract}
Single electron tunneling into small superconducting islands is sensitive to
the gap energy
of the excitations created in the process and, hence, depends on the electron
number
parity.
At low temperature the properties of the system are 2e-periodic
in the applied gate voltage, turning e-periodic at higher temperature.
We evaluate the tunneling rates and determine the probabilities for the even
and odd state
as well as the cross-over condition from the balance between different
processes.
Our analysis includes excitations in the leads. They are essential if the leads
are
superconducting.
 The influence of parity effects on single electron tunneling, Cooper pair
tunneling, and
the Andreev reflection in superconducting transistors yields rich structures
in the I-V
characteristic, which explains recent experimental findings.

 \end{abstract}
 \narrowtext

 Recent experiments showed electron number parity effects in small
superconducting
islands \cite{1,2,3}. In an electron box, if the energy gap is smaller than the
charging
energy, the voltage shows a characteristic long-short cycle, with a
$2e$-periodic
dependence on an applied gate voltage. This has been interpreted as the
even-odd
asymmetry predicted by Averin and Nazarov \cite{AN}. At higher temperatures
both
parts of the cycle become more equal in length. Eventually above a cross-over
temperature $T_{cr}$ the $e$-periodic behaviour typical for normal metal
electron
boxes is recovered.

The even-odd asymmetry arises since single electron tunneling from an initial
state with
an even number of electrons on the superconducting island leads to a state with
one extra
electron - the odd one - in an excited state. In the small island, where
charging effects
prevent further tunneling, the odd electron does not find another excitation
for
recombination. As a result the energy of the final state lies above that of the
equivalent
normal system by the gap energy. Only at larger gate
voltages another electron can enter the island, and the system relaxes to the
ground state.
The observed cross-over at higher temperatures to an $e$-periodic behavior was
explained
in Refs. \cite{1,2,3} by free energy arguments for the island. This approach
cannot be
extended to describe nonequilibrium excitations in the leads. They are
essential, however,
if the leads are superconducting and are naturally included in  the
description presented below.

We evaluate the transition rates due to electron tunneling
between a small superconducting island and normal or
superconducting lead electrodes. The rates depend on the energy difference
before
and after the tunneling, which includes the gap energy $\Delta_i$ of the
quasiparticle
excitations in the island, and $\Delta_l$ if the lead is superconducting. At
low
temperatures processes which cost energy are suppressed. This makes the
tunneling of the one excited electron, {\it the odd one}, an important process.
The rate of
this process is smaller by a factor $1/N_S$ than the total rate of the other
$N_S$
electrons in the condensate. On the other hand, the excitation energy in the
island is
regained. The balance between the rates determines
the relative probabilities of even and odd number states.

We recover the results of Ref. \cite{2} in the case of a
normal-superconducting (NS) electron box. We further provide the extension to
the
superconducting-superconducting (SS) electron box and superconducting
transistors.
The single electron tunneling current through SSS and NSN transistors shows at
low
temperature a 2e-periodic plateau structure (for suitable parameters a double
plateau).
In NSN transistors with typical parameters this current is several orders of
magnitude higher than the cotunneling current studied in Ref. \cite{AN}.
At $T > 0$  both the even and odd states are occupied. Then the Cooper pair
tunneling in
SSS  transistors and the 2e-tunneling via Andreev reflection in NSN transistors
also
connect two even or two odd states. This results in two e-shifted, 2e-periodic
sets of
current peaks, recently detected in Refs. \cite{1,herg,Kuzmin}.

At higher temperature the odd and the even state are equally populated leading
to an
$e$-periodic behaviour. The cross-over temperature $T_{cr}$ is of the order of
the gap
energy $\Delta_i + \Delta_l$, divided by the logarithm of an effective number
of
electrons. The latter is large, resulting in a substantial reduction.

We first consider an electron box \cite{2}, a closed circuit consisting of a
small
superconducting island connected via a tunnel junction with capacitance $C_J$
to a lead
electrode and via a
capacitance $C_g$ to a voltage source $U_g$. The charging energy of the system
depends
on $Q_g=C_gU_g$ and the number  $n$ of  charges on the island
\begin{equation}
E_{ch}(n, Q_g) = \frac{(ne-Q_g)^2}{2C} \; .
\label{Ech}
\end{equation}
The total capacitance is $C = C_J +C_g$. The normal state conductance
of the junction $1/R_t =4\pi e^2 N_i(0)V_iN_l(0) V_l|T|^2/\hbar$ can be
expressed by
tunneling matrix elements. The island has the normal density of states $N_i(0)$
(per unit
volume) and volume $V_i$ (similar for the lead). After a tunneling process,
starting from
an even state, one electron --- the {\it odd} one --- on the island is in an
excited state.
Therefore, the energy of the final state differs from the initial one not only
by the
charging
energy but also by the energy of the excitations $\epsilon_i \ge \Delta_i$ (and
$\epsilon_l \ge \Delta_l$ if the lead electrode is superconducting).

Since the behaviour of the system is $2e$-periodic in $Q_g$ it is sufficient to
consider
$0 \le Q_g \le 2e$. We further assume that for $0 \le Q_g \le e$ the
superconducting
island is the even state.
The transition rates due to single electron tunneling between the states with
$n=0$ and
$n=1$ electron charges (up or down) on the island are \cite{AL,5}
\begin{equation}
\Gamma^{\pm} = \frac{1}{e} I_{t}(\delta E^{\pm}(Q_g))
\frac{1}{\exp[\delta E^{\pm}_{ch}(Q_g)/T] -1} \; .
\label{rate}
\end{equation}
This is the total rate due to the tunneling  of one of the many electrons in
the junction. It
depends on the difference in the charging energy $\delta E^{\pm}_{ch}(Q_g) =
\pm
[E_{ch}(1, Q_g) - E_{ch}(0, Q_g)]$. The minimum energy
difference is $\delta E^\pm = \pm \delta E_{ch}(Q_g) + \Delta_{i} +
\Delta_l$. The energy of the excitations created in the tunneling process enter
through the
density of states into the well known classical quasiparticle tunneling
characteristic
$I_{t}(eV)$  \cite{Mike}.
 The asymptotic form of the tunneling rates for $T \ll \Delta_i$ in the SN
junctions is
\begin{equation}
\Gamma(\delta E) =
\left\{
\begin{array}{ll}
\frac{\sqrt{\pi \Delta_i T/2}}{e^2R_t} \exp \big{(} -\frac{\delta E}{T} \big{)}
&
\mbox{ for\ } \delta E > 0\\
\frac{1}{e^2R_t}\sqrt{\delta E ( \delta E - 2\Delta_i)}	& \mbox{ for\ } \delta
E < 0,
\end{array}
\right.
\label{GammaSN}
\end{equation}
For SS junctions we have for  $|\delta E| >T$ near the onset
  \begin{equation}
\Gamma(\delta E) =
\left\{
\begin{array}{ll}
\frac{\pi(\Delta_i \Delta_l)^{1/2}}{2e^2R_t}
\exp\big{(}- \frac{\delta E}{T}  \big{)}
& \mbox{ for\ }  \delta E > 0\\
\frac{\pi(\Delta_i \Delta_l)^{1/2}}{2e^2R_t}
& \mbox{ for\ }  \delta E < 0
\end{array}
\right.
\label{GammaSS}
\end{equation}

The expressions (\ref{rate}-\ref{GammaSS}) describes single electron tunneling
from a
state with even electron number to one with an odd number (more precisely from
an initial
state, where no electron is in an excited state). On the other hand, if we
consider the
tunneling from an odd electron number state (where one electron is in an
excited state)
we have to take into account two types of transitions:

\noindent
   (i) One of the many electrons in the condensate of the island can tunnel
back. The rate is
also given by eq. (\ref{rate}). The minimum energy difference is $ \delta
E^{-}(Q_g)=
-\delta E_{ch}(Q_g) + \Delta_i + \Delta_l$. Also in this transition we create
two
excitations, one on the island and one in the electrode. Hence $\delta
E^{-}(Q_g)$ is
positive in the interesting range of $Q_g$, and at low temperature the
transition rate is
exponentially small. After such a process the two excitations in the island can
recombine
fast and return the island to the ground state. The extra excitation in the
electrode will diffuse rapidly further into the leads and recombine there.
Nevertheless, energy was needed to create the two excitations, which enters
into the
transition rate.

\noindent
   (ii) In the odd state there exists a second channel: the odd electron can
tunnel back.
Its initial energy $\epsilon_i$ is distributed thermally with the constraint
$\int_0^\infty d\epsilon_i {\cal N}_i(\epsilon_i) \tilde{f}(\epsilon_i) = 1$.
The tunneling rate of the odd electron into any one of the states in the lead
is
\begin{eqnarray}
\gamma(Q_g) = 2\pi |T|^2 N_l(0)V_l\int_0^\infty d\epsilon_i
\int_{-\infty}^\infty
d\epsilon_l {\cal N}_i(\epsilon_i) {\cal N}_l(\epsilon_l)
\nonumber \\
\tilde{f}(\epsilon_i)[1-f(\epsilon_l)] \delta(\epsilon_l-\epsilon_i - \delta
E_{ch}) \; .
\label{oddrate}
\end{eqnarray}
Here, ${\cal N}_{i/l}(\epsilon) = \Theta(|\epsilon|-\Delta_{i/l})
|\epsilon|/\sqrt{\epsilon^2-\Delta_{i/l}^2}$ are the reduced BCS densities of
states. The
rate is substantially different when the odd electron can tunnel from the
lowest energy
state
$\epsilon_i = \Delta_i$ and when it has to be in a higher excited state (with
exponentially
low probability). Accordingly,
\begin{equation}
\gamma(Q_g) =
\left\{
\begin{array}{ll}
\frac{1}{2e^2R_tN_i(0)V_i}
\frac{\Delta_i + \delta E_{ch}}{\sqrt{(\Delta_i + \delta E_{ch})^2 -
\Delta_l^2} }	&
\mbox{ for\ } \delta E_{ch} -\Delta_l + \Delta_i > T\\
\frac{\sqrt{(\Delta_l/\Delta_i)}}{2e^2R_tN_i(0)V_i}
\frac{\Delta_l - \delta E_{ch}}{\sqrt{(\Delta_l - \delta E_{ch})^2 -
\Delta_i^2} }
\exp\big{(}-\frac{\Delta_l- \Delta_i-\delta E_{ch}}{T}\big{)}  & \mbox{ for\ }
\Delta_l -
\Delta_i - \delta E_{ch} >T
\end{array}
\right.
\label{deltaGamma}
\end{equation}
The rate $\gamma$ is reduced by a
factor $1/N_i(0)V_i$ compared to the transition rate  $\Gamma$ (\ref{rate}). On
the other
hand,
$\gamma$ may still be larger than $\Gamma$ since it depends on the difference
of the
gaps
$\delta E = \Delta_l - \Delta_i - \delta E_{ch}$, whereas the sum of the gaps
enters into
$\Gamma$.
The three transitions between different states are visualized in the energy
scheme in Fig.
1.  Three states appear to play a role, but since the state with 2 excitations
relaxes
quickly to the ground state by quasiparticle recombination, effectively only
one even state
and the odd state need to be considered.
However, the transition rate  $\Gamma^{oe}$ from the odd to the even state is
the sum of
the two relevant rates
\begin{equation}
\Gamma^{eo}(Q_g) =  \Gamma[\delta E^{+} ] \; , \;
\Gamma^{oe}(Q_g) =  \Gamma[\delta E^{-}]  + \gamma	\; .
\label{rates}
\end{equation}

The master equation for the occupation probabilities of these states $W_e(Q_g)$
and
$W_o(Q_g)$ is
$ d W_e(Q_g)/dt = - \Gamma^{eo}(Q_g) W_e(Q_g) + \Gamma^{oe}(Q_g)W_o(Q_g)$
with $W_e(Q_g)+W_o(Q_g) = 1$. The equilibrium solutions are
\begin{equation}
W_{e(o)}(Q_g) =\Gamma^{oe(eo)}(Q_g)/\Gamma_\Sigma(Q_g)  ,
\label{equilibrium}
\end{equation}
where $\Gamma_\Sigma(Q_g)  =\Gamma^{oe}(Q_g)+\Gamma^{eo}(Q_g)$.
For $\Gamma^{oe} \gg \Gamma^{eo}$ we have $W_e(Q_g) \approx 1$, i.e. the system
occupies the even state, while for $\Gamma^{eo} \gg \Gamma^{oe}$ the island
is in the odd state. The cross-over occurs at $W_e \approx W_o$, i.e.
\begin{equation}
\Gamma^{oe}( Q_{cr}) \approx \Gamma^{eo}( Q_{cr}).
\label{crossover}
\end{equation}

At low temperatures the odd electron tunneling
rate $\gamma$ is large compared to $\Gamma[\delta E^{-}(Q_g)]$. In this regime
we find
the even-odd asymmetric, $2e$-periodic behaviour. At high temperature $\gamma$
can be
neglected
 and the cross-over between even and
odd states occurs at the symmetry points $ Q_{cr} = e/2, 3e/2, ...$, leading to
an
$e$-periodic behaviour in $Q_g$.  The cross-over temperature $T_{cr}$ between
both
regimes can be found from the condition $\gamma(e/2) \approx \Gamma[\delta
E^{+}(
e/2)])$. Up to numerical coefficients inside the logarithm we find for $T \ll
\Delta_i$
\begin{equation}
T_{cr} =
\left\{
\begin{array}{ll}
\frac{\Delta_i+ \Delta_l}{\ln N(T_{cr})} & \mbox{ for\ } \Delta_l <  \Delta_i\\
\frac{2 \Delta_i}{\ln N_S}	& \mbox{ for\ }\Delta_i < \Delta_l
\end{array}
\right.
\label{Tcr}
\end{equation}
where $ N(T) = N_0(T)$ for $\Delta_l \ll T \ll \Delta_i$ and $N(T) = N_S$
otherwise.
 The numbers $N_0(T)= N_i(0) V_i\sqrt{2\pi\Delta_i T}$ and $N_S = \pi N_i(0)
V_i
\Delta_i$ are the number of states available for quasiparticles near the gap
and the
effective number of superconducting electrons in the island, respectively.

The second half of the parameter range $e \le Q_g \le 2e$ can be treated
analogously.
The tunneling now connects the states $n=2$ and 1. The symmetry
implies
$\Gamma^{eo/oe}(Q_g) = \Gamma^{eo/oe}(2e-Q_g)$.
This leads to the following picture for $T < T_{cr}$:
For $- Q_{cr} < Q_g < Q_{cr}(T) < e$ the system is in the even state, then it
switches to the
odd state where it stays for $Q_{cr} < Q_g < 2e - Q_{cr}$, and periodic beyond.

The switching point is temperature dependent. For $\Delta_l < \Delta_i$ we find
for $ T < T_l \equiv 2\Delta_l/\ln N_S$
 \begin{equation}
Q_{cr}(T) =
 \frac{C\Delta_i}{e} + \frac{e}{2} - \frac{CT}{2e} \ln N(T)
\label{Qcr1}
 \end{equation}
 whereas for $T_l < T < T_{cr}$
 \begin{equation}
 Q_{cr}(T) = \frac{C}{e}(\Delta_i + \Delta_l) + \frac{e}{2} - \frac{CT}{e} \ln
N(T) \: .
\label{Qcr2}
\end{equation}
For $\Delta_l > \Delta_i$ the result (\ref{Qcr1}) with $N(T) = N_S$ is valid at
any
temperature $T < T_{cr}$.

The expressions for $Q_{cr}$ were derived for $\Delta_i < E_C \equiv e^2/2C$.
In
the opposite case, $\Delta_i > E_C$, for $T < T_0$ tunneling of 2 electrons
becomes
more favourable and the two-state model
is not sufficient anymore. Here we defined
\begin{equation}
T_0 =
\left\{
\begin{array}{ll}
(\Delta_i +\Delta_l -E_C)/ \ln N(T)	& \mbox{ for\ } \Delta_l < \Delta_i-E_C\\
2(\Delta_i -E_C)/ \ln N_S	& \mbox{ for\ } \Delta_l > \Delta_i-E_C
\end{array}
\right.
\label{T0}
\end{equation}
On the other hand, for $T > T_0$ the results given above remain
valid.

In the case where the  lead electrode is normal $\Delta_l = 0$ we reproduce the
results of
Refs. \cite{1,2,3}, which were derived from considerations of the free energy
of the
superconducting island only. If the electrode is
superconducting the energy of excitations in the lead $\epsilon_l \ge
\Delta_l$ needs to be taken into account, since the transition rates depend on
the total
energy difference. In contrast to the excitation which is trapped in the
island, the
excitations in the lead diffuse away  on a time scale fast in comparison to
inverse
tunneling rates. Hence, their energy cannot be regained. This behavior
involving different
time scales is not
accounted for in an equilibrium free energy description.

The analysis presented above can also be extended to describe even-odd effects
in NSN and
SSS transistors. In this system the charging energy depends also on the
transport voltage
and on the number of electrons transported through the transistor. The total
capacitance
of the island $C = C_g + C_l + C_r$ defines $E_C = e^2/2C$. The energy
differences for
tunneling in the left and right junctions are
\begin{equation}
\delta E_{ch,l/r} = E_C - \frac{e(Q_g \pm Q_{tr}/2)}{C}
\label{Echs}
\end{equation}
For definiteness we assume  $C_l=C_r$ and $eU_{tr} < 2E_C$, and we define
$Q_{tr} = C
U_{tr}$. It is again sufficient to consider only one even and one odd state.
The transitions
in this system are described by a master equation with the rates
\begin{equation}
\Gamma^{eo} \rightarrow \Gamma^{eo}_l + \Gamma^{eo}_r
\label{sssgamma}
\end{equation}
which depend on the energy differences $\delta E_l$ and  $\delta E_r$. They
again are the
sum of $\delta E_{ch,l/r}$ and the energies of the excitations created in the
island and
electrodes. A similar relation holds for the transition from the odd state to
the even state.
After the substitution (\ref{sssgamma}) the stationary solutions for the
occupation
probabilities are given by eq. (\ref{equilibrium}), and the cross-over gate
voltage
$\{Q_g\}_{cr}\equiv Q_{cr}$ follows from (\ref{crossover}).
It depends on the transport voltage, which opens the possibility to tune the
cross-over
condition. The effect becomes visible only for $T < eU_{tr}$.  Then the two
dominant
tunneling processes are tunneling in the left junction with rate
$\Gamma^{eo}_l$ followed
by a tunneling process in the right junction $\Gamma^{oe}_r$. Both balance each
other in
equilibrium. The cross-over temperature is defined by the condition  $Q_{cr}(T)
\approx
e/2$.
For the SSS transistor (with $\Delta_l=\Delta_i$) it is
\begin{equation}
T_{cr} = [2\Delta-eU_{tr}/2]/\ln N_S \; .
\label{sssTcr}
\end{equation}

Next we evaluate the current due to single electron tunneling through the NSN
or SSS
transistors. For high temperatures $T>T_{cr}$ the current is
$I = e[ \Gamma^{eo}_l \Gamma^{oe}_r - \Gamma^{eo}_r
\Gamma^{oe}_l]/\Gamma_\Sigma$,
where the rates are given in (\ref{rate}). It
has maxima at the points $Q_g =e/2 + ne$.
At lower temperature  $T<T_{cr}$  the current is $I=e\gamma_r$. It is
observable in the
window
 $Q_{cr}(T) < Q_g < \frac{e}{2} + \Delta C/e + Q_{tr}/2$ and exponentially
small outside. In
NSN transistors it reduces to
\begin{equation}
I = 1/[2eR_tN_i(0) V_i]
\end{equation}
 A second current peak or plateau exists in the window
$3e/2 -\Delta C/e - Q_{tr}/2 < Q_g <2e - Q_{cr}$. Both plateaus create a double
structure
which repeats 2e-periodically. For $\Delta + eU_{tr}/2 > E_C$ these plateaus
merge
forming a 2e-periodic single plateau structure. Note that the current (19) is
much larger
(by a factor $10^2$ for parameters of ref. [6]) than the cotunneling current
[4]. It can
explain the presence of a constant current $80 fA$ of "unknown origin" detected
in Ref.
\cite{herg}.

The occupation probabilities $W_e$ and $W_o$, which are regulated by the single
electron
tunneling processes, also influence the supercurrent through SSS transistors
and the
Andreev reflection in NSN systems. If the SSS transistors is in the even state
Cooper
pairs can tunnel through the system at $Q_g = \pm e, \pm 3e, ... $ for small
transport
voltages. This leads to a set of 2e-periodic sharp resonant peaks in the I-V
curves.
The amplitude of this peak is reduced below the $T=0$ result $I_e = I(T=0)
W_e(e)$.
Analogously, if the system is in the
odd state Cooper pair tunneling through the system occurs at $Q_g = 0, \pm 2e,
... $, with
current peak amplitude $I_o = I(T=0) W_o(0)$. These probabilities are
\begin{eqnarray}
W_e(e) = \frac{e^{-2\Delta/T}+1/N_S}{e^{-2(\Delta-E_C)/T}+1/N_S}
\nonumber \\
W_o(0) = \frac{e^{-(2\Delta+E_C)/T}}{e^{-(2\Delta-E_C)/T}+1/N_S} \;.
\label{CooperpeakN}
\end{eqnarray}
At $T>T^*=2\Delta/\ln(N_S)$ both current peaks are equal in height and the
behaviour of
the system is e-periodic. At lower temperature the two peaks are distinct. A
behaviour of
this type has recently been observed in experiments \cite{1,Kuzmin}, the
$T$-dependence
is in qualitative agreement with the results of Ref \cite{Kuzmin}.

If we consider NSN transistors we have to distinguish two cases. For $\Delta <
E_C$ the
single electron tunneling is crucial as discussed above. In the other limit
$\Delta > E_C$
the mechanism of Andreev reflection transferring 2 electrons becomes important
\cite{GS,7}.
Close to $Q_g = \pm e, \pm 3e, ... $ at $T=0$ the shape of the current
resonance
\begin{equation}
I_{res}(\delta U_g, U_{tr}) = G(\delta U_g,U_{tr})
\Big(U_{tr} - \frac{4C_g^2}{U_{tr}C^2}(\delta U_g)^2\Big)
\label{Hekking}
\end{equation}
has been calculated in Ref. \cite{7} for tunneling between even states.  At
finite
temperature, due to single
electron tunneling, there exists a finite probability for the odd state. This
leads to an
additional set of peaks due to 2e tunneling between two odd states. The
amplitudes of the
even-even and odd-odd current resonances then are
$I_{ee} = I_{res} W_e(U_g-2E_C,U_{tr}) $ for $Q_g
\approx e$ and $I_{oo} = I_{res} W_o(U_g,U_{tr})$ for $Q_g \approx 0$,
where
\begin{equation}
W_e(Q_g\approx e) = \frac{e^{-(\Delta+E^-_C)/T}+1/[N_0(T)\cosh\frac{e\delta
Q_g}{CT}]}{e^{-(\Delta-E^-_C)/T}+1/[N_0(T)\cosh\frac{e\delta Q_g}{CT}]} \; ,
\label{NSN2}
\end{equation}
\begin{equation}
W_o(Q_g\approx 0) = \frac{e^{-(\Delta+E_C-e|U_g|)/T}}
{e^{-(\Delta-E_C+e|U_g|)/T}+1/[N_0(T)\cosh\frac{eU_{tr}}{2T}]}
\label{Cooperpeak}
\end{equation}
Here $\delta Q_g = Q_g - e, E^{ \pm }_C = E_C \pm eU_{tr}/2$. The two e-shifted
peaks
acquire equal height and the picture becomes e-periodic above a
cross-over temperature $T_{A}^* = [\Delta + E^-_C ]/\ln N_0(T) \ge T_{cr}$. The
presence
of the odd peaks $I_{oo}$  has been clearly demonstrated in
recent experiments \cite{herg}.

For metals with a short elastic mean free path $l$ at low temperatures and
transport
voltages the effective conductance $G$ in eq. (\ref{Hekking}) near the maximum
of the
current is \cite{ADZ}
\begin{equation}
G(0,U_{tr}) =\frac{12d}{e^2 lR_t^2 p_F^2 S}
\frac{\Delta^2}{\Delta^2 - (E^+_C)^2}
\Big{[}\arctan{\sqrt{\frac{\Delta+E^+_C}{\Delta-E^+_C}}}  \Big{]}^2 \; .
\label{theG}
\end{equation}
Here S is the junction area, $d$ the size of the normal leads which have
resistance  small
compared to $R_t$. In contrast to the result quoted in Ref. \cite{HN}, the
conductance
(\ref{theG}) remains finite at $T=eU_{tr}=0$.  For larger temperature or
voltage
$\delta = \max(T, eU_{tr}/2) > E_d = v_F l/3d^2 $ the conductance $G$ is
smaller than
(\ref{theG}) by a factor of order $(E_d/ \delta)^{1/2}$ and roughly agrees with
the result
of Ref. \cite{HN} in this limit. The result (\ref{theG}) is consistent with the
height of the
experimentally observed current peaks \cite{3,herg}.

In conclusion, we have developed a theory of parity effects in small
superconducting
islands by analyzing the rates of electron tunneling. Nonequilibrium
excitations  in the
superconducting leads generated by the tunneling processes turn out to be
essential. Our
theory explains a large number of recent experiments in the electron box or
transistors
and can easily be extended to more general situations.

We would like to thank M. Devoret, D. Esteve, F.W. Hekking, J.M. Hergenrother,
M. Tinkham,
and Yu.V. Nazarov for stimulating discussions.
  This work is part of ``Sonderforschungsbereich 195'' supported by Deutsche
 Forschungsgemeinschaft. We also acknowledge the support by
 a NATO Linkage Grant.

 
\vspace{1cm}

\begin{large}
\noindent
{\bf Figure Captions :}

\end{large}

\vspace{1cm}

\noindent
Fig. 1 :
Energy of different states in the NS electron box. Shown are the regular single
electron
tunneling transitions $\Gamma^\pm$ and the transition $\gamma$ due to the
tunneling of
the odd electron. The state with two excitations and energy $E > 2\Delta$
relaxes quickly
to the ground state by quasiparticle recombination.
\end{document}